\documentclass{article}

\usepackage[preprint]{neurips_2024}

\usepackage[table]{xcolor}
\usepackage{bbm}
\usepackage[utf8]{inputenc} 
\usepackage[T1]{fontenc}    
\usepackage{url}            
\usepackage{booktabs}       
\usepackage{amsfonts}       
\usepackage{nicefrac}       
\usepackage{microtype}      
\usepackage{xcolor}         
\usepackage{graphicx}
\usepackage{amsmath}
\usepackage{amssymb}
\usepackage{multirow} 
\usepackage{subcaption}
\usepackage{enumitem}
\usepackage{capt-of}
\usepackage{hyperref}

\definecolor{LightGray}{gray}{0.92}

\title{YingMusic-SVC: Real-World Robust Zero-Shot Singing Voice Conversion with Flow-GRPO and Singing-Specific Inductive Biases}

\author{
\textbf{Gongyu Chen}$^{1,}$\thanks{Equal contribution.} \quad
\textbf{Xiaoyu Zhang}$^{1,2,*}$ \quad
Zhenqiang Weng$^{1,3}$ \quad
Junjie Zheng$^{1}$ \\
\textbf{Da Shen}$^{4}$ \quad
\textbf{Chaofan Ding}$^{1}$ \quad
\textbf{Wei-Qiang Zhang}$^{4}$ \quad
\textbf{Zihao Chen}$^{1}$ \\[4pt]
$^{1}$AI Lab, GiantNetwork  \quad
$^{2}$University College London (UCL) \\
$^{3}$East China University of Science and Technology \quad
$^{4}$SATLab, Tsinghua University \\[3pt]
}

\begin{document}

\maketitle
\setcounter{footnote}{0}

\begin{abstract}
Singing voice conversion (SVC) aims to render the target singer’s timbre while preserving melody and lyrics. However, existing zero-shot SVC systems remain fragile in real songs due to harmony interference, F0 errors, and the lack of inductive biases for singing.
We propose YingMusic-SVC, a robust zero-shot framework that unifies continuous pre-training, robust supervised fine-tuning, and Flow-GRPO reinforcement learning. Our model introduces a singing-trained RVC timbre shifter for timbre–content disentanglement, an F0-aware timbre adaptor for dynamic vocal expression, and an energy-balanced rectified flow matching loss to enhance high-frequency fidelity.
Experiments on a graded multi-track benchmark show that YingMusic-SVC achieves consistent improvements over strong open-source baselines in timbre similarity, intelligibility, and perceptual naturalness—especially under accompanied and harmony-contaminated conditions—demonstrating its effectiveness for real-world SVC deployment. Our code and models are available at:~\url{https://github.com/GiantAILab/YingMusic-SVC}
\end{abstract}

\begin{center}
    \centering
    \includegraphics[width=0.99\linewidth]{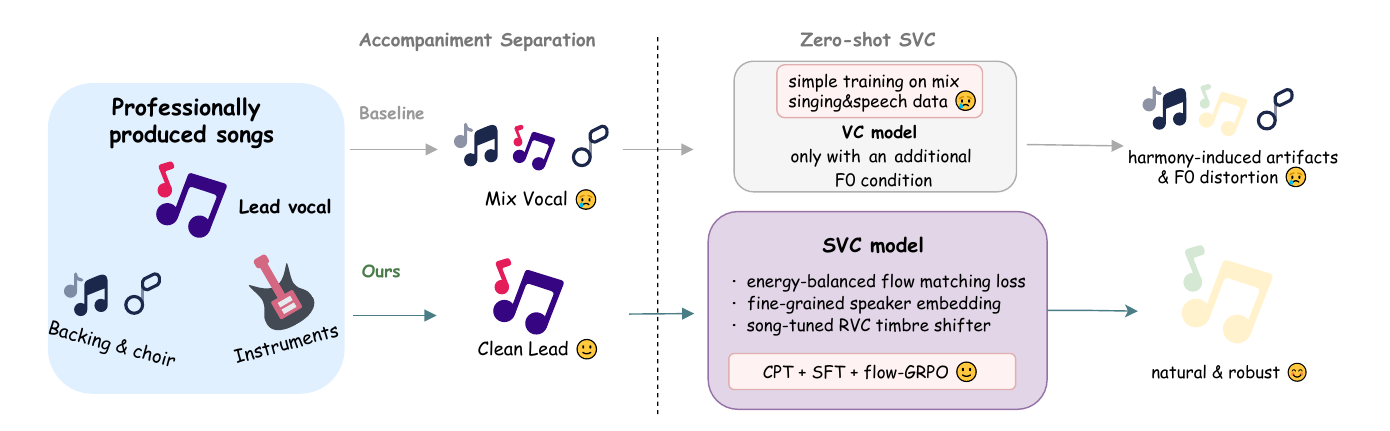}
    \captionof{figure}{From professionally produced songs to zero-shot SVC: baseline limitations and our improved pipeline with singing-oriented designs.}
    \label{fig:svc_head}
\end{center}

\section{Introduction}



Singing voice conversion (SVC)~\citep{spasvc, 11210176, chen2019singing, mohammadi2017overview, engel2019gansynth} aims to transform the vocal timbre of a source singer into that of a target singer while preserving the original musical content. This technique has promising applications in creative music production, virtual singers, and social media content creation. Recently, any-to-many SVC works such as So-VITS-SVC~\citep{sovits2023} and RVC~\citep{rvc2023} have been able to achieve realistic conversion effects, but they lack robustness and require fine-tuning using target speaker data; The zero-shot SVC model~\citep{liu2024zero} has achieved good conversion results on the pure vocal test set under laboratory conditions. However, there remains a substantial gap between current research prototypes and the requirements of real-world industrial applications. On one hand, most SVC systems are evaluated on ideal clean vocals and struggle when faced with full songs that include background accompaniment. In practical pipelines, due to the scarcity of available a cappella vocal tracks, a music source separation module is typically applied upfront to isolate vocals from the mixed audio~\citep{10800547, 10887782}. This two-stage approach introduces new challenges: contemporary music productions often add backing vocals or harmony layers to the lead vocal~\citep{hennequin2020spleeter}, which means the separated “vocals” track may still contain residual artifacts. Converting such imperfect vocals can lead to audible artifacts in the output; furthermore, any inaccuracies in fundamental frequency (F0) extraction~\citep{kim2018crepe, Wei_2023} under noisy conditions can result in off-key or distorted converted singing. On the other hand, most zero-shot voice conversion approaches simply incorporate an F0 conditioning on top of a speech VC architecture, without imparting inductive biases tailored for singing. These generic designs fail to account for key characteristics of singing voices – for example, singing vocals exhibit much larger dynamics and timbral variability, and their spectral content contains richer high-frequency harmonic components than normal speech~\citep{sundberg1987science,lundy2000acoustic,  blaauw2017neural, gupta2022deep}. Consequently, vanilla zero-shot VC models often underperform on singing data, as they do not sufficiently emphasize the high-frequency details and other singing-specific attributes. As illustrated in Figure \ref{fig:svc_head}, these limitations frequently manifest as harmony leakage, F0 instability, and spectral artifacts when current zero-shot SVC models are applied to professionally produced songs.

To bridge these gaps, we propose a comprehensive set of improvements for robust zero-shot SVC in industrial scenarios. First, regarding pipeline robustness, we enhance the vocal separation front-end using internal multi-track data, yielding cleaner vocal stems free of harmony bleed. In parallel, we devise a robust fine-tuning (SFT) training strategy for the SVC model: by augmenting training with lightly distorted and harmony-contaminated vocals, we improve the model’s resilience to real-world noisy inputs, thereby mitigating artifacts like voice cracking and residual harmony in the conversions. Second, in terms of model design, we introduce novel modules and loss functions tailored to singing audio. We incorporate an F0-aware fine-grained timbre adaptation module that allows the model to dynamically adjust timbre features conditioned on pitch, enhancing the expressive dynamics of the converted singing. Furthermore, in order to neither lose high-frequency information nor leak timbre features during content feature extraction, we adopted the RVC model trained with singing data as the timbre shifter. We also design an energy-balanced flow matching loss~\citep{lipman2022flow} that rebalances the training objective to give more emphasis to high-frequency, low-energy regions of the spectrum, encouraging the model to pay extra attention to fine high-frequency details of the singing voice. Finally, we explore a reinforcement learning (RL)~\citep{sutton1998reinforcement} based post-training paradigm for SVC by applying a Flow-GRPO algorithm~\citep{liu2025flow} to refine the model. We formulate a multi-objective reward that combines speech quality metrics with aesthetic style measures, thereby simultaneously improving the converted singing’s technical quality and artistic expressiveness.

Our main contributions are summarized as follows:
\begin{itemize}
    
\item \textbf{An end-to-end robust SVC pipeline for industrial use.} We develop an improved accompaniment separation model and a robustness-enhanced SVC training process, which together significantly improve conversion quality on songs with accompaniment, reducing artifacts from residual harmonies and F0 extraction errors. A benchmark and difficulty classification test set for real scenarios was also designed and a comprehensive assessment was conducted.

\item \textbf{Singing-specific modeling enhancements.} We introduce an F0-aware fine-grained timbre adaptation module, a song-trained RVC timbre shifter and an energy-balanced flow matching loss, which substantially improve the model’s ability to capture the dynamic timbre range and high-frequency details of singing voices, leading to more natural and faithful conversions.

\item \textbf{RL fine-tuning and open-source implementation for SVC.} We present the first application of reinforcement learning in the Diffusion Transformer (DiT)~\citep{peebles2023scalable} based SVC, designing a multi-reward function that jointly optimizes vocal quality and aesthetic expression in the converted singing. We also open-source a full industry-grade SVC pipeline implementation, to spur further research and application.
\end{itemize}
The experimental results show that this method is comprehensively superior to the baseline on our graded real-world singing test set. Our system produces significantly higher-quality and more natural and robust converted singing, especially in the presence of background music, confirming the effectiveness of the introduced techniques.

\section{Related Work}
\subsection{Singing Voice Conversion (SVC)}

Open-source SVC frameworks such as So-VITS-SVC and RVC have become strong community baselines. So-VITS-SVC combines a HuBERT-based content encoder~\citep{hsu2021hubert} with a VITS-style generative model~\citep{kim2021conditional} to separate linguistic and speaker information for high-fidelity conversion. RVC extends this pipeline with a feature retrieval module that injects nearest-neighbor embeddings from a training set to better preserve target timbre. These systems are widely used in practice and form the foundation for many one-shot and any-to-many SVC applications.

Recent work has advanced zero-shot SVC, enabling conversion to unseen singers. LDM-SVC~\citep{chen2024ldm} uses a latent diffusion model and classifier-free guidance to suppress source leakage and improve conversion quality, though at high computational cost. SaMoye~\citep{wang2024samoye} disentangles content, pitch, and timbre features, and strengthens speaker modeling using augmented embeddings from similar singers, achieving robust performance even in cross-domain tasks. R2-SVC~\citep{zheng2025r2} improves real-world robustness through data augmentation and singing-informed encoders, effectively handling noisy and accompanied inputs. HQ-SVC~\citep{bai2025hq} adopts a unified codec and introduces an EVA module for better timbre-pitch alignment, along with diffusion-based refinement. These models highlight various strategies—disentanglement, latent modeling, or enhanced supervision—to improve generalization and quality in zero-shot settings. However, many existing SVC systems either lack robustness under real-world mixtures and noisy conditions, or struggle to capture the large dynamic timbre range and rich high-frequency details of singing voices. These limitations motivate our work on a more robust, singing-oriented zero-shot SVC pipeline.

\subsection{Reinforcement Learning in Speech and Generative Modeling}

In speech and audio generation, supervised objectives based on spectral reconstruction or phoneme accuracy improve intelligibility but fail to capture perceptual attributes such as naturalness, stability, timbral consistency, and musicality—key qualities for high-fidelity SVC. Since these attributes lack reliable labels, supervised training often diverges from human perception, motivating the adoption of RL to directly optimize non-differentiable perceptual metrics and align models with human preferences.

Recent work has begun to introduce reinforcement learning and preference alignment into speech generation. INTP~\citep{zhang2025advancing} extends Direct Preference Optimization (DPO)~\citep{rafailov2023direct} to Text-to-Speech (TTS) with large-scale intelligibility preference data, while F5R-TTS~\citep{sun2025f5r} adapts Group Relative Policy Optimization (GRPO)~\citep{shao2024deepseekmathpushinglimitsmathematical} to flow-matching TTS with rewards on word error rate and speaker similarity, improving zero-shot voice cloning performance. In broader generative modeling, Flow-GRPO~\citep{liu2025flow} proposes an online RL framework for flow-matching text-to-image models via ODE–SDE conversion and KL-regularized GRPO updates. However, these methods focus on TTS or image synthesis; to our knowledge, there is still little work applying online RL to zero-shot SVC, particularly for DiT-based flow-matching architectures and multi-objective rewards that jointly target intelligibility, timbre similarity, and musical aesthetics under real-world singing conditions.

\section{Approach}
\label{sec:approach}

\subsection{Overview}
We focus on the singing voice conversion (SVC) task, which involves transforming an input singing voice into the voice of a target singer while preserving the musical content (melody and lyrics). Formally, given an input audio (or its acoustic features) $x$ from a source singer and a target singer identity $s_{\text{tgt}}$, the goal is to generate an output audio $y$ that carries the timbre of $s_{\text{tgt}}$ but sings the same content as $x$. We adopt a diffusion-based generative framework built upon the Seed-VC architecture (a state-of-the-art DiT model for VC). The model operates on mel-spectrograms and uses a rectified flow matching objective to learn a continuous velocity field $v_\theta(x(t), t)$ that drives a differential generative process. In essence, $v_\theta$ learns to transform a noise distribution at $t=1$ into the target mel-spectrogram at $t=0$.


\begin{figure*}[t!]
    \centering
    \includegraphics[width=0.99\linewidth]{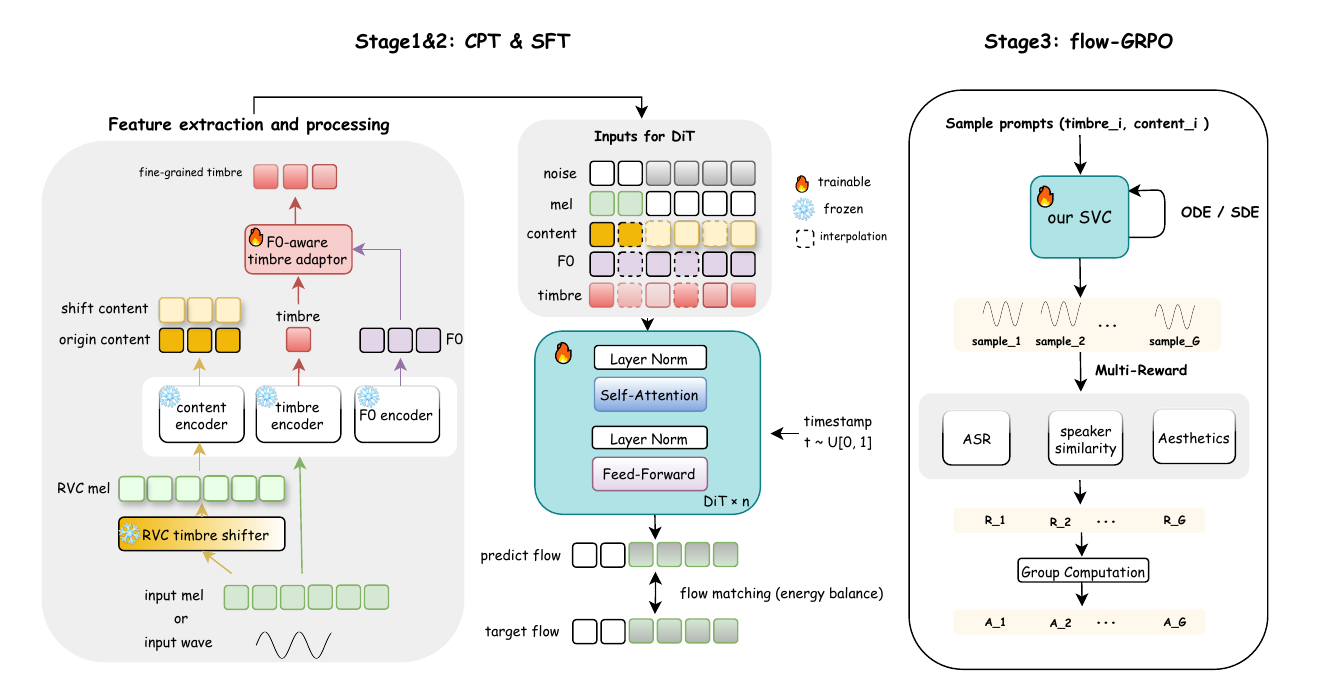}
    \caption{Three-Stage Training Framework of the Proposed YingMusic-SVC Model.}
    \label{fig:svc_main}
\end{figure*}

Beginning from a Seed-VC checkpoint, we adopt a three-stage training regime to obtain a robust zero-shot SVC model: a continuous pre-training (CPT) stage to adapt and stabilize the newly introduced singing-specific modules, a supervised fine-tuning (SFT) stage on curated and augmented singing corpora to enhance robustness, and a RL post-training stage based on Flow-GRPO with multi-objective rewards that directly optimize perceptual and aesthetic attributes of the converted singing. As illustrated in Figure~\ref{fig:svc_main}, the data flow in the first two stages is unified. Given an input vocal $x$, we first compute its log-mel spectrogram $m=\mathcal{M}(x)$. Three frozen encoders are then applied to extract different representations from $m$:
\begin{equation}
h_c^{\text{orig}}=\mathcal{E}_c(m), \quad 
e_\tau=\mathcal{E}_\tau(x), \quad 
h_f=\mathcal{E}_f(m),
\end{equation}
where $\mathcal{E}_c$, $\mathcal{E}_\tau$, and $\mathcal{E}_f$ denote the content, timbre, and F0 encoders, respectively. 
To suppress timbre leakage from the source, we employ a pretrained RVC model $\mathcal{T}_{\text{RVC}}$ to perform random-target SVC on $x$, obtaining a shifted version $\tilde{x}=\mathcal{T}_{\text{RVC}}(x; s_{\text{rand}})$. After shifting the timbre-related information, its mel is re-encoded to produce a \textit{timbre-shifted content} feature:
\begin{equation}
h_c^{\text{shift}}=\mathcal{E}_c(\mathcal{M}(\tilde{x})).
\end{equation}
This feature is used as the main content representation for generation, while $h_c^{\text{orig}}$ can optionally serve for auxiliary reconstruction consistency.

The global timbre embedding $e_\tau$ and pitch-related feature $h_f$ are fused through a lightweight F0-aware timbre adaptor $\mathcal{A}_{\text{F0}}$, yielding a fine-grained timbre map aligned with the local melodic contour:
\begin{equation}
h_\tau=\mathcal{A}_{\text{F0}}(e_\tau, h_f).
\end{equation}

\paragraph{Temporal Assembly for DiT.}
Following the simulated-inference strategy in Seed-VC, we construct time-dependent conditional inputs with a stochastic masking scheme, a random boundary $t_m \sim \mathcal{U}(0,1)$ is sampled for each training iteration. The first $t_m T$ frames are dominated by timbre conditioning, while the remaining frames provide content and pitch cues:
\begin{equation}
m_{\tau}(t)=\mathbbm{1}[t/T < t_m], \quad 
m_{c}(t)=\mathbbm{1}[t/T \ge t_m].
\end{equation}
All features are temporally aligned to the mel frame length via nearest-neighbor interpolation:
\begin{equation}
\widehat{h}_c^x=\text{NNInterp}(h_c^x, T), \quad
\widehat{h}_f=\text{NNInterp}(h_f, T), \quad
\widehat{h}_\tau=\text{NNInterp}(h_\tau, T).
\end{equation}
where $x\in \{orig, shift\}$.

During conditional assembly, the content stream is split into two temporal parts following the randomly sampled mask $m_c(t)$:
the first segment (observed region) uses the original content $\widehat{h}_c^{\text{orig}}$, while the second segment (prediction region) uses the shifted content $\widehat{h}_c^{\text{shift}}$ to suppress source-timbre bias.
Formally, the masked content feature is defined as
\begin{equation}
\widetilde{h}_c(t) = m_{\tau}(t)\,\widehat{h}_c^{\text{orig}}(t) + m_c(t)\,\widehat{h}_c^{\text{shift}}(t).
\end{equation}
The DiT conditional input at time $t$ is then concatenated along the feature dimension as
\begin{equation}
z_{\text{cond}}(t) =
[\widehat{h}_\tau(t),\,
\widetilde{h}_c(t),\,
\widehat{h}_f(t)].
\end{equation}

For mel modeling, we also apply the same temporal mask to the input mel:
\begin{equation}
\tilde m(t) = (1-m_c(t))\,m(t) + m_c(t)\,x_t, 
\quad x_t=(1-t)m + t\varepsilon,\ \varepsilon\sim\mathcal{N}(0,\mathbf{I}),
\end{equation}
so that the model observes the first segment of ground-truth mel while predicting the masked future region along the rectified flow path.
The DiT-based rectified flow model $\mathbf{v}_\theta$ takes $(\tilde m, z_{\text{cond}}, t)$ as input and predicts the velocity field in mel space for the masked region.
During training, the flow-matching loss is applied to the latter segment of frames (content-provided region), with energy-balanced weighting to emphasize low-energy, high-frequency components crucial for singing timbre reconstruction.
This conditioning pathway is shared across CPT, SFT, and RL, while the optimization objective transitions from pure rectified flow matching in CPT/SFT to a combination of flow-matching and RL-derived advantages in the Flow-GRPO stage.

\subsection{Singing-Specific Modeling Enhancements}

In addition to the training strategy, we introduce several model-level improvements to the Seed-VC backbone to better handle singing voice characteristics:

\paragraph{RVC Timbre Shifter for Content Extraction.} Instead of directly feeding the raw input vocals to the diffusion model, we incorporate a Retrieval-based Voice Conversion (RVC) module as a timbre shifter. This module, pre-trained on a diverse set of 120 singers, converts the input singing voice into a random auxiliary timbre before content encoding. Formally, let $x$ be the separated vocal input and let $s_{\text{rand}}$ be a randomly chosen speaker ID from the RVC’s training set. We generate $\tilde{x}=\mathcal{T}_{\text{RVC}}(x; s_{\text{rand}})$, an audio where the content (melody and phonetics) of $x$ is preserved but the timbre is shifted to a different singer. This procedure encourages the downstream content encoder to focus on linguistic and musical content while discarding the source speaker identity. We then extract content features $h_c=\mathcal{E}_c(m)$ using a content encoder $\mathcal{E}_c$ (e.g., a pre-trained ASR encoder). By disrupting the original timbre, the content features become more speaker-invariant and better suited for conversion to the target voice. In practice, replacing Seed-VC’s original timbre shifter with the RVC-based one led to more accurate pronunciations and melody retention in the outputs, since the model receives cleaner content features unaffected by source timbre idiosyncrasies.

\begin{figure*}[t!]
    \centering
    \includegraphics[width=0.99\linewidth]{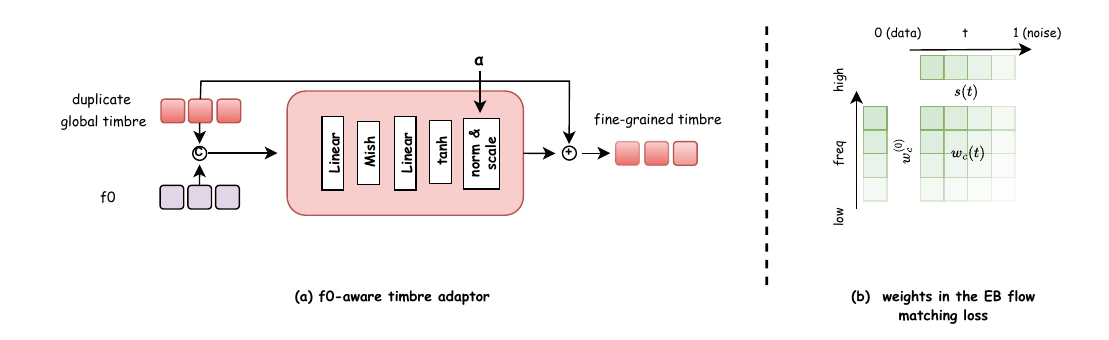}
    \caption{Key components of YingMusic-SVC.
(a) F0-aware timbre adaptor that refines global timbre embeddings into fine-grained, pitch-sensitive representations.
(b) Energy-balance flow matching loss with time- and frequency-dependent weighting to enhance high-frequency reconstruction fidelity.}
    \label{fig:method}
\end{figure*}

\paragraph{F0-Aware Adaptive Speaker Embedding.} Singing voices exhibit strong dynamic timbre variations across pitches and expressions (e.g., belting in high notes vs. soft low notes). To model this, we design the speaker conditioning to be time-varying and fundamental-frequency-aware rather than a static embedding, similar to recent style adaptor approaches in SVC~\citep{style_adaptor_svc}. As shown in Figure \ref{fig:method}(a), We decompose the target singer’s representation into a static global vector and a time-varying residual that depends on F0. Let ${e_\tau}^{(global)} \in \mathbb{R}^d$ be the learned global speaker embedding for the target singer, and $h_{f}(t)$ be an embedding of the input’s F0 at time $t$. We feed both into a small network (an MLP) to predict a timbre shift residual $\Delta e_{\tau}(t)$: 
\begin{equation}
\Delta e_{\tau}(t) = \text{MLP}\Big(\big[h_{f}(t);\; e_{\tau}^{(global)}\big]\Big)
\end{equation}
where $[u;v]$ denotes concatenation of vectors $u$ and $v$. The final speaker conditioning at time $t$ is then $h_{\tau}(t) = e_{\tau}^{(global)} + \alpha_{\tau}\cdot\Delta e_{\tau}(t)$, where $\alpha_{\tau}$ is the hyperparameter used to adjust the strength of the residual.
 This $h_{\tau}(t)$ is injected into the diffusion model in place of the usual static embedding. Intuitively, $\Delta e_\tau(t)$ allows the model to modulate the timbral color depending on the current pitch: for example, adding brightness or head resonance at higher pitches, or more chest resonance at lower pitches, emulating how a real singer’s voice quality changes with pitch. By providing F0 context to the model’s speaker representation, we observed improvements in capturing expressiveness and avoiding timbre distortions on extreme notes.

\paragraph{Frequency-Aware Energy-Balanced Loss.} We modify the flow matching loss to address the imbalance in energy across frequency bands in singing audio, inspired by ~\citep{rfwave}. High-frequency component often has much lower energy than lower frequencies on the mel-spectrum, which can lead a plain MSE loss to under-weight high-frequency reconstruction. To ensure better high-frequency modeling, we introduce a frequency-dependent weighting scheme
in the loss. Let $\mathbf{u}(t) \in \mathbb{R}^C$ denote the ground-truth velocity field at time $t$ for each frequency channel $c$ (e.g., each mel frequency bin), and $\mathbf{v}_\theta(t)$ the model’s predicted velocity. The original loss is $L_{\text{flow}} = \mathbb{E}_{t}\big[|\mathbf{u}(t) - \mathbf{v}_\theta(t)|^2\big]$. We introduce a weight vector $\mathbf{w}(t) = [w_1(t), ..., w_C(t)]$ to emphasize certain components:
\begin{equation}
L_{\text{flow}}^{\text{EB}} = \mathbb{E}_{t}\Big[ \sum_{c=1}^C w_c(t)\, \big(u_c(t) - v_{\theta,c}(t)\big)^2 \Big]
\end{equation}
 where $w_c(t)$ is defined to allocate higher importance to low-energy (especially high-frequency) channels and at later diffusion steps. As shown in figure \ref{fig:method}(b), we construct $w_c(t)$ as follows. First, we estimate the inverse amplitude of each channel: $w_c^{(0)} = 1/\sigma_c$, where $\sigma_c$ is the variance of $u_c(t)$ over time and samples on the training data. This $w_c^{(0)}$ acts as a energy-inverse weight, making channels with weaker energy (smaller $\sigma_c$) receive larger weight. Next, we define a monotonically increasing function $g(c)$ that boosts higher frequency indices (for instance, $g(c)$ linearly increases from $0$ for low $c$ to $1$ for the highest $30\%$ of frequencies). We also define a time-dependent factor $s(t)$ that increases as $t$ approaches 0 (end of the diffusion) to emphasize detail in later denoising stages; specifically $s(t) = (1-t)^2 / \mathbb{E}[(1-t)^2]$ so that $\mathbb{E}[s(t)] = 1$ (here $t\sim U[0,1]$). Finally, we combine these to obtain:
$w_c(t) = w_c^{(0)} \Big(1 + \lambda \, s(t)\, g(c)\Big), $
 where $\lambda$ is a hyperparameter controlling the strength of the frequency-time emphasis. In practice, we also normalize $\mathbf{w}(t)$ for each sample to have mean 1 (so the overall scale of the loss remains similar). This energy-balanced loss gives more weight to the higher frequencies especially towards the end of diffusion (when fine details are generated), while not overly penalizing early stages or low frequencies. By using this scheme, the model learns to generate crisp high-frequency details without sacrificing stability.

\subsection{Robust SFT for Real-World Singing}

The supervised fine-tuning (SFT) stage preserves the overall data flow and conditioning structure established during CPT, but introduces additional robustness-oriented augmentations to address artifacts frequently encountered in real-world singing. In practical SVC pipelines, vocals obtained from accompaniment separation often contain residual harmony or backing layers, and pitch estimation may accidentally lock onto harmonic components rather than the true fundamental. These issues can lead to unstable timbre, octave errors, and audible pitch breaks. To mitigate such effects, we incorporate two complementary strategies: stochastic perturbation of F0 features inspired by R2-SVC~\citep{zheng2025r2} and harmony-contaminated multi-track augmentation.

\paragraph{F0 perturbation.}  
Let $h_f$ denote the pitch embedding extracted from the frozen F0 encoder. During SFT, we apply random perturbations that simulate realistic F0 noise:
\begin{equation}
h_f^{\text{aug}} = h_f + \Delta_f,
\end{equation}
where $\Delta_f$ is sampled from a mixture of perturbation kernels that emulate jitter, sliding, and abrupt jumps:
\begin{equation}
\Delta_f \sim 
\begin{cases}
\text{Jitter}(\sigma_j), & p = p_j,\\
\text{Glide}(\ell_g), & p = p_g,\\
\text{Jump}(\delta_k), & p = p_k.
\end{cases}
\end{equation}
This forces the model to remain stable under noisy pitch trajectories and prevents over-reliance on exact F0 estimates.

\paragraph{Multi-track harmony augmentation.}  
For content- and mel-based branches, we simulate imperfect separation by mixing backing tracks into the lead vocal. Let $x_{\text{lead}}$ and $x_{\text{harm}}$ denote the lead and harmony tracks. The augmented input becomes
\begin{equation}
x_{\text{mix}} = x_{\text{lead}} + \alpha\, x_{\text{harm}},
\end{equation}
where $\alpha$ controls the contamination strength. The model is still supervised against the clean ground-truth mel 
$m_{\text{lead}}$, using the mixed input only for robustness-oriented
conditioning. We construct the conditional representation as:
\begin{equation}
z_{\text{cond}}^{\text{mix}}(t)=
\big[
\widehat{h}_\tau^{\text{mix}}(t),\;
\widetilde{h}_c^{\text{mix}}(t),\;
\widehat{h}_f^{\text{aug}}(t)
\big],
\end{equation}
where $\widehat{h}_\tau^{\text{mix}}(t)$ and 
$\widetilde{h}_c^{\text{mix}}(t)$ are timbre/content representations extracted 
from $m_{\text{mix}}$, and $\widehat{h}_f^{\text{aug}}(t)$ is the interpolated (time-aligned) form of
the perturbed F0 embedding $h_f^{\text{aug}}$.

During SFT, the DiT-based flow matching model is trained to map corrupted
inputs toward the clean vocal trajectory, forcing the model to suppress harmony leakage and remain stable under pitch jitter.

\textbf{Combined, the final augmented conditioning pipeline is summarized as:}
\begin{equation}
\mathbf{v}_\theta\big(\tilde{m}_{\text{mix}}, z_{\text{cond}}^{\text{mix}}(t), t\big)
\;\longrightarrow\;
\text{flow}(m_{\text{lead}}),
\end{equation}
which highlights that robustness is jointly achieved via 
\emph{F0 perturbation} and \emph{multi-track harmony augmentation}, enabling
the model to handle residual backing vocals and unstable pitch estimates in 
real-world SVC scenarios.

\subsection{Reinforcement Learning with Flow-GRPO}
\label{sec:grpo}

In the final stage, we refine the model through online reinforcement learning to directly optimize non-differentiable perceptual objectives. Following Flow-GRPO~\citep{liu2025flow}, we reinterpret the rectified flow model~\citep{liu2022flow} as a stochastic policy by converting the deterministic ODE trajectory into an SDE formulation:
\begin{equation}
d x_t =\left(v_\theta(x_t, t)- \frac{\sigma_t^2}{2}\,\nabla \log p_t(x_t)\right) dt + \sigma_t\, d w_t,
\end{equation}
which admits an Euler--Maruyama discretization~\citep{kloeden1977numerical} that yields an explicit Markovian transition $\pi_\theta(x_{t-1}\mid x_t)$ suitable for KL-regularized policy-gradient optimization. Here, \(d w_t\) denotes increments of a standard Wiener process, and \(\sigma_t\) is a  
time-dependent diffusion coefficient controlling the level of stochasticity injected at time \(t\).  
We parameterize the noise schedule as
\begin{equation}
\sigma_t = a \sqrt{\frac{t}{1 - t}}, 
\end{equation}
where \(a\) acts as a scalar coefficient that determines the overall intensity of the injected noise.

Under this formulation, the policy update maximizes:
\begin{equation}
J(\theta)= \mathbb{E}_{\pi_\theta}\!\left[R(x_0,c) - \beta\, D_{\mathrm{KL}}(\pi_\theta\,\|\,\pi_{\mathrm{ref}})\right],
\end{equation}
where $R$ is a perceptual reward and $\pi_{\mathrm{ref}}$ is a frozen reference policy.

\vspace{2pt}
\noindent\textbf{Stochastic Timestep Selection for SDE.}  
Direct SDE sampling introduces credit-assignment ambiguity across the full trajectory~\citep{team2025longcat}. To mitigate the temporal credit-assignment ambiguity introduced by full-trajectory SDE sampling, we adopt a selective-noise strategy that injects randomness at a single timestep, which has become a common design choice in recent flow-based RL work~\citep{team2025longcat}.
 For each conditioning prompt $c$, all samples in a group share the same initial noise, and stochasticity is introduced only at one uniformly sampled timestep $t'\sim\mathcal{U}(0,T'-1)$; all other steps follow deterministic ODE updates. This design isolates reward differences to a single transition, enabling more interpretable and stable gradients.  
The resulting training objective becomes:
\begin{equation}
\scalebox{0.85}{$
\begin{aligned}
J_{\text{GRPO-Selective}}(\theta)
&=\mathbb{E}_{c,\;t',\;\{x^i\}_{i=1}^G \sim \pi_{\mathrm{old}}}\!
\left[\frac{1}{G}\sum_{i=1}^{G}
\Big(r_{t'}^i(\theta)\, \hat{A}^i - 
\beta\, D_{\mathrm{KL}}\!\left(\pi_\theta\,\|\,\pi_{\mathrm{ref}}\right)_{t'}\Big)
\right],
\end{aligned}
$}
\end{equation}
where $r_{t'}^i$ is the policy ratio correcting for off-policy sampling, and $\hat{A}^i$ denotes the group-relative advantage. This selective-noise scheme preserves exploration while substantially improving optimization stability.

\vspace{6pt}
\noindent\textbf{Reward models.}  
To align the model with perceptual and musical goals specific to SVC, we employ three reward families:

\begin{itemize}[leftmargin=10pt]
    \item \textbf{Aesthetic quality.}  
    We use Meta's Audiobox Aesthetics model\footnote{https://github.com/facebookresearch/audiobox-aesthetics} ~\citep{tjandra2025meta}and select the perceptual dimensions \emph{Content Enjoyment (CE)} and \emph{Content Usefulness (CU)}, defining:
    \begin{equation}
        R_{\mathrm{aes}}=\frac{R_{\mathrm{CE}}+R_{\mathrm{CU}}}{2}.
    \end{equation}

    \item \textbf{Lyrics intelligibility.}  
    To evaluate whether the converted singing preserves clear articulation and linguistic content, we employ an ASR-based metric~\citep{xu2025fireredasr} and compute:
    \begin{equation}
        R_{\mathrm{int}} = 1 - \mathrm{WER}.
    \end{equation}

    \item \textbf{Timbre similarity.}  
    Speaker-level consistency is measured via a pretrained embedding model, using cosine similarity between the generated and reference timbre embeddings~\citep{resemblyzer}:
    \begin{equation}
        R_{\mathrm{spk}} = \cos(e_{\mathrm{gen}}, e_{\mathrm{ref}}).
    \end{equation}
\end{itemize}

Before integration, all reward values are normalized to $[0,1]$. The final multi-objective reward for sample $i$ is:
\begin{equation}
R^i = \sum_k w_k R_k^i,\quad w_k=1,
\end{equation}
and the group-relative advantage is:
\begin{equation}
\hat{A}^i = \frac{R^i - \mu}{\sigma},
\end{equation}
where $\mu$ and $\sigma$ denote the mean and standard deviation over the group for the same conditioning prompt.  
This multi-objective RL framework enables direct optimization of perceptual quality, intelligibility, and timbral fidelity, complementing the supervised objectives in preceding stages.

\subsection{Inference Procedure}

In practical industrial SVC pipelines, the input audio is typically a full song containing both vocals and accompaniment. To provide the SVC model with a clean vocal signal and avoid artifacts caused by residual background components, we employ a dedicated music source separation front-end. Specifically, we use a customized Band RoFormer separator~\citep{wang2024mel} trained in-house on approximately 3600 multi-track songs (about 250 hours), optimized to decompose complex musical mixtures into disentangled components.  

Formally, given a mixture waveform $x_{\mathrm{mix}}$, the separator $\mathcal{D}_{\mathrm{BR}}$ produces:
\begin{equation}
(\,x_{\mathrm{lead}},\; x_{\mathrm{back}},\; x_{\mathrm{inst}}\,)
= 
\mathcal{D}_{\mathrm{BR}}(x_{\mathrm{mix}}),
\end{equation}
where $x_{\mathrm{lead}}$ denotes the lead vocal, $x_{\mathrm{back}}$ the backing/harmony vocals, and $x_{\mathrm{inst}}$ the instrumental accompaniment.  
Our optimized Band RoFormer design follows a band-split Transformer architecture operating on multi-band spectrogram partitions, and is trained with adversarial objectives using both Multi-Scale and Multi-Period discriminators. This setup enhances the perceptual realism and reduces leakage between sources.

For SVC inference, only the separated lead vocal $x_{\mathrm{lead}}$ is forwarded through our conversion pipeline. The backing vocal estimate $x_{\mathrm{back}}$ may optionally be retained for downstream mixing or post-production, depending on the application. After conversion, the predicted target-singer mel-spectrogram is converted back to waveform via the vocoder, and can be recombined with $x_{\mathrm{inst}}$ to reconstruct a complete song. Thus, the full inference chain can be summarized as:
\begin{equation}
x_{\mathrm{tgt}} 
= \mathrm{Vocoder}\!\left(
\mathrm{SVC}(x_{\mathrm{lead}}, s_{\mathrm{tgt}})
\right),
\qquad
x_{\mathrm{final}} 
= x_{\mathrm{tgt}} + \gamma_\mathrm{inst} x_{\mathrm{inst}}.
\end{equation}
where $\gamma_\mathrm{inst}$ is a hyperparameter used to control the relative loudness of the accompaniment, which defaults to 1.

This separation–conversion–recomposition workflow ensures that inference remains robust under real-world recording and mixing conditions, and provides an end-to-end pipeline suitable for industrial SVC deployment.
For cross-gender conversion scenarios, we apply a pitch transposition of $\pm12$\,semitones after F0 feature extraction. This normalization step is widely adopted in SVC systems, and prevents octave errors or instability when the source and target singers exhibit large differences in vocal range.

\section{Experiments}
\label{sec:exp}

\subsection{Dataset}
Our training data comprise three complementary sources.
(1) Speech corpus: we sample 1,000 hours of English and 1,000 hours of Chinese speech from the Emilia dataset \citep{emilia} to provide rich linguistic coverage.
(2) Open singing datasets: six public singing corpora—GTSinger \citep{zhang2024gtsinger}, M4Singer \citep{zhang2022m4singer}, OpenCpop \citep{wang2022opencpop}, OpenSinger \citep{huang2021multi}, PopBuTFy \citep{liu2022learning}, and PopCS \citep{liu2022diffsinger}—are combined, totaling roughly 230 hours of clean solo vocals.
(3) Internal multi-track dataset: $500$ hours of professionally produced recordings containing aligned lead and backing vocal stems, which provide realistic mixtures and timbre diversity.

We use different subsets across training stages: CPT utilizes the main (lead) tracks from all three sources (1 + 2 + 3), SFT is performed on the clean and multi-track singing data (2 + 3), and RL fine-tuning is conducted on the open clean singing data (2) to ensure generalization. 
To ensure compatibility with reward computation, we additionally filter out samples that are not in Chinese or English, remove pure-speech recordings unrelated to singing, and discard clips shorter than $5$ seconds that tend to produce unstable or degenerate intelligibility rewards.

For evaluation, we construct a difficulty-graded test suite to emulate real production conditions. Using the internal multi-track data, we randomly sample 100 10–20 s segments containing semantically meaningful singing content. Among them, 20 converted samples are randomly selected for subjective evaluation, which is conducted by a panel of 10 human listeners. The simple test set (GT Leading) includes only the lead vocal track, while the complex test set (Mix Vocal) merges the lead and backing tracks to simulate vocals obtained from source-separation front-ends. In addition, we evaluate a third setting (Ours Leading), where the lead vocals are obtained from our in-house separation model, allowing us to assess end-to-end robustness within our actual production pipeline. Five target timbres, covering both genders and multiple age ranges, are selected for conversion. All evaluation speakers and songs are unseen during training.


\subsection{Implementation Details}
YingMusic-SVC is built upon an open-source SVC implementation, Seed-VC \citep{liu2024zero}, which utilizes a DiT-based flow matching architecture. The model consists of 17 DiT blocks with 12 attention heads, an embedding dimension of 768, and an FFN dimension of 3072. The audio is processed at a sampling rate of 44.1 kHz, using a 2048-point FFT window, a hop size of 512, and 128 mel-frequency bins. Models were trained with an effective batch size of 80, consisting of 60,000 CPT steps and 15,000 SFT steps before the RL stage. We use the AdamW optimizer with a peak learning rate of $1\times10^{-4}$, which exponentially decays to a minimum of $1\times10^{-5}$. During RL post-training, we update the model using full-parameter optimization. The model is initialized from the supervised checkpoint prior to RL updates. Table~\ref{tab:rl_settings} summarizes the main hyperparameters used in our RL training. Finally, we employ the pretrained BigVGAN\footnote{https://huggingface.co/nvidia/bigvgan\_v2\_44khz\_128band\_512x} \citep{lee2022bigvgan} model as the vocoder.


For random $F_{0}$ perturbations, we set the probabilities of the three perturbation types to
$p_{\text{j}} = 0.1$, $p_{\text{g}} = 0.1$, and $p_{\text{k}} = 0.3$.
Each audio sample contains between 2 and 4 perturbed $F_{0}$ segments.
In the F0-aware timbre adaptor, the interpolation coefficient is fixed to $\alpha_{\tau} = 0.5$, balancing global timbre embedding and local pitch-dependent modulation.
For the energy-balanced loss, we set the frequency-time emphasis strength to $\lambda = 0.4$, which provides sufficient weighting of high-frequency, low-energy components without destabilizing early-stage flow updates.


\begin{table}[t]
\centering
\caption{Main hyperparameter settings for RL training.}
\label{tab:rl_settings}
\begin{tabular}{lclc}
\toprule
\textbf{Parameter} & \textbf{Value} & \textbf{Parameter} & \textbf{Value} \\
\midrule
Group size & 8 & Prompts per step & 8\\
Noise level $a$ & 0.4 & SDE steps range & [0, 6]\\
Learning rate & \(1\times 10^{-5}\) & Iterations & 0.8k\\
\# Sampling steps & 10 &  & \\
\bottomrule
\end{tabular}
\end{table}

\subsection{Baselines}
We compare YingMusic-SVC with two other systems: Seed-VC\footnote{https://github.com/Plachtaa/seed-vc} ~\citep{liu2024zero}and FreeSVC\footnote{https://github.com/freds0/free-svc}~\citep{10890068}. As one of the state-of-the-art open-source voice conversion systems, we use the official 200M-parameter Seed-VC checkpoint\footnote{https://huggingface.co/Plachta/Seed-VC/tree/main}, which has been specifically trained for singing voice conversion.

\subsection{Evaluation Metrics}
For objective evaluation, we examine four dimensions: speaker similarity, intelligibility, pitch consistency, and aesthetic quality. Speaker similarity is measured using cosine similarity between speaker embeddings (SPK-SIM), computed with Resemblyzer\footnote{https://github.com/resemble-ai/Resemblyzer}
. Intelligibility is evaluated using character error rate (CER) obtained from FireRedASR\footnote{https://github.com/FireRedTeam/FireRedASR} , an ASR model trained on data containing both speech and singing. To assess pitch accuracy and stability, we report the log-F0 Pearson correlation coefficient (LogF0PCC) between the converted and reference F0 contours. Aesthetic quality is quantified using the Unified Automatic Quality Assessment model from Meta’s Audiobox Aesthetics\footnote{https://github.com/facebookresearch/audiobox-aesthetics}.
For subjective evaluation, we conduct Mean Opinion Score (MOS) listening tests, measuring perceptual naturalness (NMOS) and speaker similarity (SMOS) based on human listener ratings.

\section{Results}
\subsection{Overall Performance Across Difficulty-Graded Test Sets}

Table~\ref{tab:svc_metrics} summarizes the objective and subjective results across three evaluation settings designed to reflect increasing levels of real-world degradation. On the \textbf{GT Leading} set, where clean vocal tracks are used, our models consistently outperform FreeSVC and Seed-VC across all metrics. In particular, \textit{Ours-SFT} achieves the highest CMOS (4.37) and SMOS (3.15), indicating substantial perceptual gains over Seed-VC (3.98/3.09), while maintaining comparable pitch consistency (98.08\% vs. 98.29\%). 
The \textit{Ours-Full} model further improves aesthetic quality, attaining the best CE/CU scores (5.86/6.56).

Under the more challenging \textbf{Mix Vocal} condition—where lead and backing vocals are mixed to emulate source-separation artifacts—the performance gap becomes more pronounced. Seed-VC suffers significant degradation in intelligibility (CER increases from 10.89\% to 17.30\%) and pitch stability (LogF0PCC drops from 98.29\% to 84.02\%). In contrast, our robustness-enhanced models maintain strong performance: \textit{Ours-Full} achieves the best aesthetic scores (5.75/6.40), the highest CMOS (3.31), and the highest pitch consistency (86.47\%), outperforming Seed-VC by a large margin across all metrics.

Finally, the \textbf{Ours Leading} set evaluates the full industrial pipeline using vocals separated by our in-house Band-Roformer front-end. Despite the additional real-world noise factors, our models continue to outperform both baselines. \textit{Ours-Full} delivers the best CE/CU (5.73/6.39), the highest CMOS (3.91), and the strongest timbre similarity (0.801), demonstrating that the proposed system maintains robustness even when evaluated end-to-end under production conditions.

Overall, these results show that our method consistently improves timbre similarity, intelligibility, pitch accuracy, and perceptual quality across all difficulty levels, with the largest gains observed in the most challenging real-world scenarios.

\begin{table}[t]
\centering
\small
\renewcommand{\arraystretch}{0.85}
\setlength{\tabcolsep}{7pt}
\caption{Objective and subjective results on three SVC test settings (GT Leading, Mix Vocal, Ours Leading).
Best in \textbf{bold}; $\uparrow$/$\downarrow$ denote higher/lower is better.}
\label{tab:svc_metrics}
\resizebox{\textwidth}{!}{%
\begin{tabular}{lccccccc}
\toprule
& \multicolumn{5}{c}{\textbf{Obj. Metrics}} & \multicolumn{2}{c}{\textbf{Sub. Metrics}}\\
\cmidrule(lr){2-6} \cmidrule(lr){7-8}
 &
\multirow{2}{*}{\textbf{SPK-SIM}$\uparrow$} &
\multirow{2}{*}{\textbf{CER(\%)}$\downarrow$} &
\multicolumn{2}{c}{\textbf{Aesthetics}} &
\multirow{2}{*}{\textbf{LogF0PCC(\%)}$\uparrow$} &
\multirow{2}{*}{\textbf{CMOS}$\uparrow$} &
\multirow{2}{*}{\textbf{SMOS}$\uparrow$} \\
\textbf{Model} & & & \textbf{CE}$\uparrow$ & \textbf{CU}$\uparrow$ & & & \\
\midrule
\multicolumn{8}{l}{\textbf{Setting 1: GT Leading}}\\
\cmidrule(lr){1-8}
FreeSVC~\citep{10890068}    & 0.648 & 20.70 & 4.72 & 5.55 & 86.56 & 2.47  & 1.83     \\
Seed-VC~\citep{liu2024zero} & 0.801 & 10.89 & 5.69 & 6.39 & \textbf{98.29} & 3.98 & 3.09 \\
\midrule
\rowcolor{LightGray} \textit{Ours-CPT}                 & 0.802 & \textbf{9.20} & 5.79 & 6.50 & 98.27 & 4.15 & 3.02 \\
\rowcolor{LightGray} \textit{Ours-SFT}           & \textbf{0.804} & 9.66 & 5.84 & 6.52 & 98.08 & \textbf{4.37} & \textbf{3.15} \\
\rowcolor{LightGray} \textit{Ours-Full}        & 0.800 & 9.26 & \textbf{5.86} & \textbf{6.56} & 98.12 & 4.26  & 3.08     \\
\midrule
\multicolumn{8}{l}{\textbf{Setting 2: Mix Vocal}}\\
\cmidrule(lr){1-8}
FreeSVC~\citep{10890068}    & 0.657 & 33.26 & 4.07 & 5.13 & 60.18 & 2.02     & 1.91     \\
Seed-VC~\citep{liu2024zero} & 0.808 & 17.30 & 5.51 & 6.17 & 84.02 & 2.93 & 2.78 \\
\midrule
\rowcolor{LightGray}
\textit{Ours-CPT}                 & \textbf{0.810} & 15.70 & 5.65 & 6.32 & 84.29 & 2.85 & 2.92 \\
\rowcolor{LightGray}
\textit{Ours-SFT}           & 0.809 & \textbf{15.40} & 5.73 & 6.35 & 86.12 & 3.23 & 2.95 \\
\rowcolor{LightGray}
\textit{Ours-Full}        & 0.805 & 15.90 & \textbf{5.75} & \textbf{6.40} & \textbf{86.47} & \textbf{3.31}     & \textbf{3.12}     \\
\midrule
\multicolumn{8}{l}{\textbf{Setting 3: Ours Leading}}\\
\cmidrule(lr){1-8}
FreeSVC~\citep{10890068}    & 0.650 & 33.94 & 4.36 & 5.32 & 75.64 & 2.31   & 1.90     \\
Seed-VC~\citep{liu2024zero} & 0.800 & \textbf{18.64} & 5.53 & 6.21 & 89.18 & 3.46 & 3.00 \\
\midrule
\rowcolor{LightGray} \textit{Ours-CPT}         & 0.801 &  18.65 & 5.66 & 6.30 & 89.27 & 3.51 & 3.02 \\
\rowcolor{LightGray} \textit{Ours-SFT}           & \textbf{0.803} &  18.93 & 5.72 & 6.34 & 89.65 & 3.86 & \textbf{3.11} \\
\rowcolor{LightGray} \textit{Ours-Full}        & 0.801 &  18.89 & \textbf{5.73} & \textbf{6.39} & \textbf{89.78} &  \textbf{3.91}  & 3.06     \\
\bottomrule
\end{tabular}
}
\end{table}

\begin{table}[t]
\centering
\tiny
\renewcommand{\arraystretch}{0.8}
\setlength{\tabcolsep}{10pt}
\caption{
Ablation study of key singing-specific components across three SVC test settings 
(GT Leading, Mix Vocal, Ours Leading). 
Arrows ($\uparrow$ / $\downarrow$) denote higher / lower is better, respectively.
}
\label{tab:svc_ablation}
\resizebox{\textwidth}{!}{

\begin{tabular}{lccccc}
\toprule
& \multirow{2}{*}{\textbf{SPK-SIM}$\uparrow$} &
\multirow{2}{*}{\textbf{CER(\%)}$\downarrow$} &
\multicolumn{2}{c}{\textbf{Aesthetics}} &
\multirow{2}{*}{\textbf{LogF0PCC(\%)}$\uparrow$} \\
\textbf{Model} & & & \textbf{CE}$\uparrow$ & \textbf{CU}$\uparrow$ & \\
\midrule
\multicolumn{6}{l}{\textbf{Setting 1: GT Leading}}\\
\cmidrule(lr){1-6}
\rowcolor{LightGray} \textit{Ours-CPT}                 & 0.802 & 9.20 & 5.79 & 6.50 & 98.27  \\
\midrule
\textit{w/o RVC-ts}                 & 0.798 & 9.73 & 5.71 & 6.40 & 98.26  \\
\textit{w/o timbre adaptor}           & 0.803 & 9.40 & 5.75 & 6.46 & 98.17  \\
\textit{w/o EB loss}        & 0.804 & 9.40 & 5.81 & 6.54 & 98.38    \\
\midrule
\multicolumn{6}{l}{\textbf{Setting 2: Mix Vocal}}\\
\cmidrule(lr){1-6}
\rowcolor{LightGray} \textit{Ours-CPT}    & 0.810 & 15.70 & 5.65 & 6.32 & 84.29  \\
\midrule
\textit{w/o RVC-ts}                 & 0.804 & 16.30 & 5.54 & 6.18 & 84.41  \\
\textit{w/o timbre adaptor}           & 0.806 & 16.10 & 5.61 & 6.27 & 84.55  \\
\textit{w/o EB loss}      & 0.808 & 15.80 & 5.65 & 6.33 & 84.36    \\
\midrule
\multicolumn{6}{l}{\textbf{Setting 3: Ours Leading}}\\
\cmidrule(lr){1-6}
\rowcolor{LightGray} \textit{Ours-CPT}         & 0.801 &  18.65 & 5.66 & 6.30 & 89.27  \\
\midrule
\textit{w/o RVC-ts}         & 0.798 &  18.85 & 5.53 & 6.19 & 89.18  \\
\textit{w/o timbre adaptor}        & 0.801 &  18.10 & 5.60 & 6.25 & 89.01  \\
\textit{w/o EB loss}        & 0.803 &  18.95 & 5.67 & 6.36 &89.42  \\
\bottomrule
\end{tabular}
}
\end{table}

\subsection{Effect of Multi-Stage Training (CPT $\rightarrow$ SFT $\rightarrow$ RL)}

The progressive training pipeline yields clear and complementary improvements across stages. 
\textbf{CPT} establishes a solid singing-specific representation by integrating the RVC timbre shifter, the F0-aware timbre adaptor, and the energy-balanced loss. This stage already surpasses Seed-VC in nearly all metrics across all settings. For example, CPT reduces CER from 10.89\% to 9.20\% on GT Leading and improves LogF0PCC from 84.02\% to 84.29\% under Mix Vocal.

\textbf{SFT} provides noticeable robustness gains by training the model on augmented inputs containing F0 perturbations and simulated harmony leakage. This improvement becomes more observable under noisy conditions. On Mix Vocal, SFT reduces CER from 15.70\% to 15.40\%, raises LogF0PCC from 84.29\% to 86.12\%, and boosts CMOS from 2.85 to 3.23. Similar gains appear on Ours Leading, where SFT improves both CMOS (3.51 to 3.86) and SMOS (3.02 to 3.11).

Finally, \textbf{RL} post-training refines perceptual quality and timbre fidelity through multi-objective optimization. On Mix Vocal, RL further improves all aesthetic dimensions—CE increases from 5.73 to 5.75 and CU from 6.35 to 6.40—and boosts CMOS from 3.23 to 3.31. On Ours Leading, RL yields the highest CMOS (3.91), demonstrating its effectiveness in enhancing real-world perceptual attributes that are not fully captured by supervised training.

Together, these results illustrate that each training stage contributes distinct benefits: CPT provides singing-aware inductive bias, SFT delivers robustness under realistic distortions, and RL enhances higher-level perceptual preferences. The full model integrates these strengths and consistently achieves the best overall performance across evaluation settings.

\subsection{Ablation Study}
\subsubsection{Ablation of Singing-Specific Modules}

We first analyze the contribution of the three singing-specific components introduced in the CPT stage: the RVC-based timbre shifter, the F0-aware timbre adaptor, and the energy-balanced (EB) flow-matching loss. Table~\ref{tab:svc_ablation} reports ablation results across the GT Leading, Mix Vocal, and Ours Leading settings.

On the \textbf{GT Leading} set, removing the RVC timbre shifter (\textit{w/o RVC-ts}) leads to a consistent decrease in performance relative to the CPT baseline, with lower SPK-SIM (0.798 vs.\ 0.802), higher CER (9.73\% vs.\ 9.20\%), and reduced aesthetic scores (CE 5.71 vs.\ 5.79; CU 6.40 vs.\ 6.50). Removing the timbre adaptor also results in mild degradation in aesthetic metrics (5.75/6.46 vs.\ 5.79/6.50). The EB loss shows a different pattern: while CE/CU slightly increase (5.81/6.54), CER rises from 9.20\% to 9.40\%, indicating a trade-off between spectral detail and intelligibility under clean conditions.

In the more challenging \textbf{Mix Vocal} setting, the effects of each component become more pronounced. Removing the RVC timbre shifter produces the largest degradation among the three ablations, with SPK-SIM decreasing from 0.810 to 0.804, CER increasing from 15.70\% to 16.30\%, and aesthetic scores dropping to 5.54/6.18. Eliminating the timbre adaptor also lowers CE/CU (5.61/6.27) and slightly increases CER (16.10\%). The EB loss again shows a mixed effect: CE/CU remain close to the baseline (5.65/6.33), while CER slightly increases from 15.70\% to 15.80\%. Overall, all three components contribute positively to performance under acoustic interference, with the RVC-based timbre shifter yielding the largest improvement.

On the \textbf{Ours Leading} set, which reflects the full production pipeline, the trends remain consistent. Removing the RVC timbre shifter leads to a decrease in SPK-SIM (0.798 vs.\ 0.801) and aesthetic quality (5.53/6.19 vs.\ 5.66/6.30). Removing the timbre adaptor results in deterioration of CE/CU (5.60/6.25), while removing the EB loss causes CER to rise from 18.65\% to 18.95\%. Although the EB loss yields slightly higher CE/CU (5.67/6.36), its higher CER indicates less stable F0 and content reconstruction under real-world interference.

Taken together, these results demonstrate that all three modules provide measurable benefits across evaluation conditions. The RVC-based timbre shifter contributes most to timbre preservation and intelligibility in noisy settings, the timbre adaptor improves local timbre–pitch alignment, and the EB loss enhances high-frequency reconstruction while requiring balance with intelligibility. Their combined effect yields the strongest overall performance in the CPT configuration.

\begin{figure}[t]
    \centering
    \begin{subfigure}{0.24\linewidth}
        \centering
        \includegraphics[width=\linewidth]{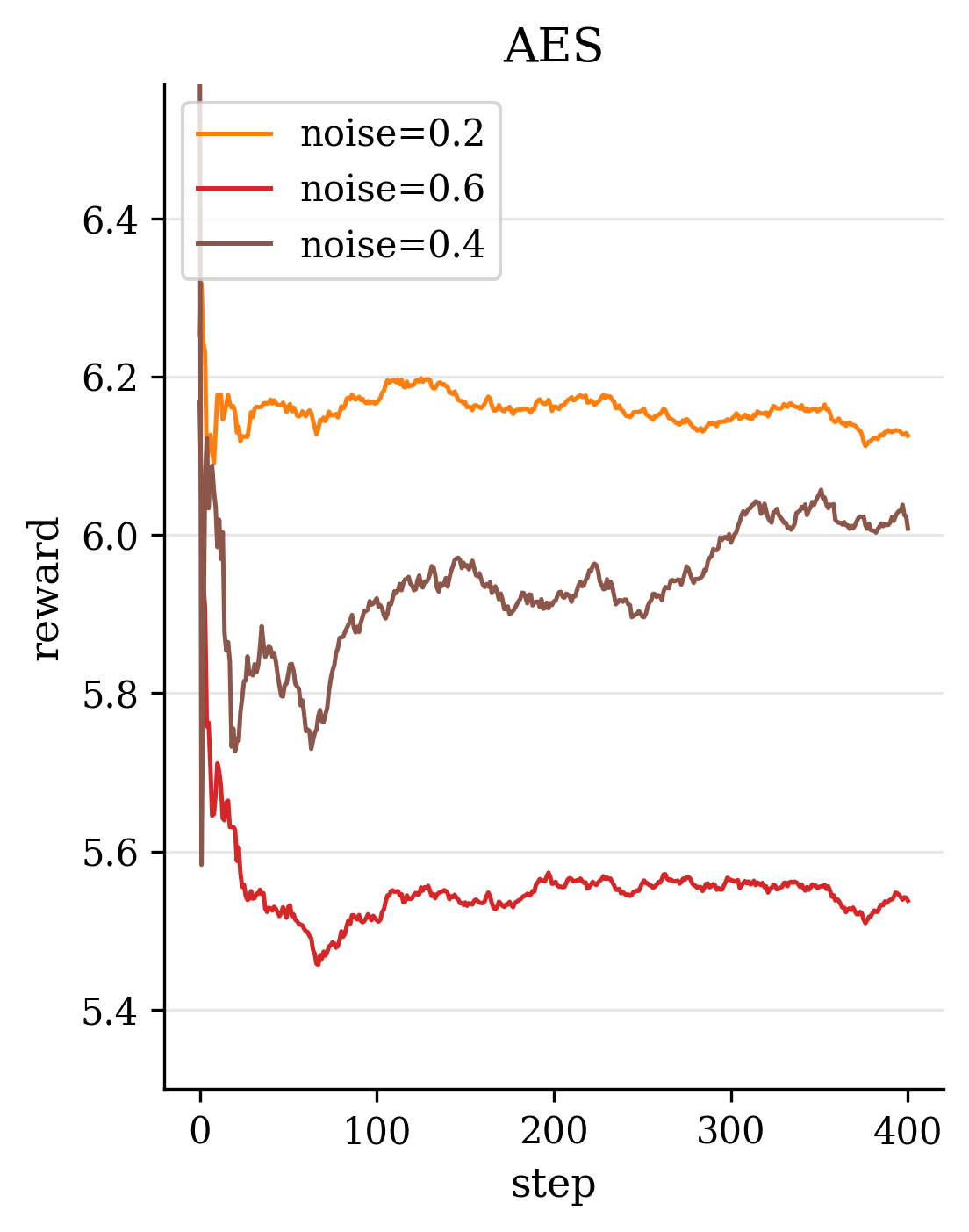}

        \label{fig:sub1}
    \end{subfigure}
    \hfill
    \begin{subfigure}{0.24\linewidth}
        \centering
        \includegraphics[width=\linewidth]{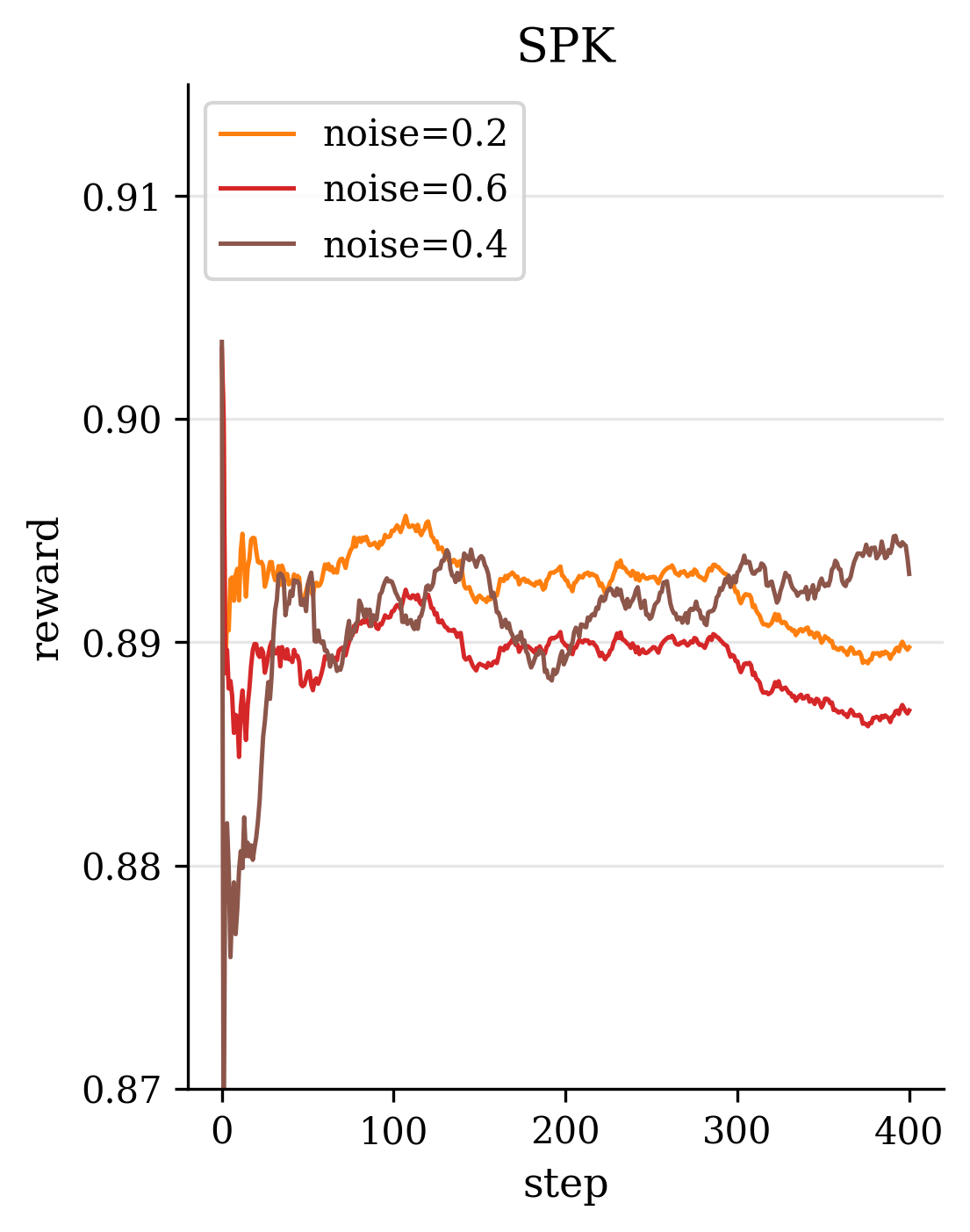}

        \label{fig:sub2}
    \end{subfigure}
    \hfill
    \begin{subfigure}{0.24\linewidth}
        \centering
        \includegraphics[width=\linewidth]{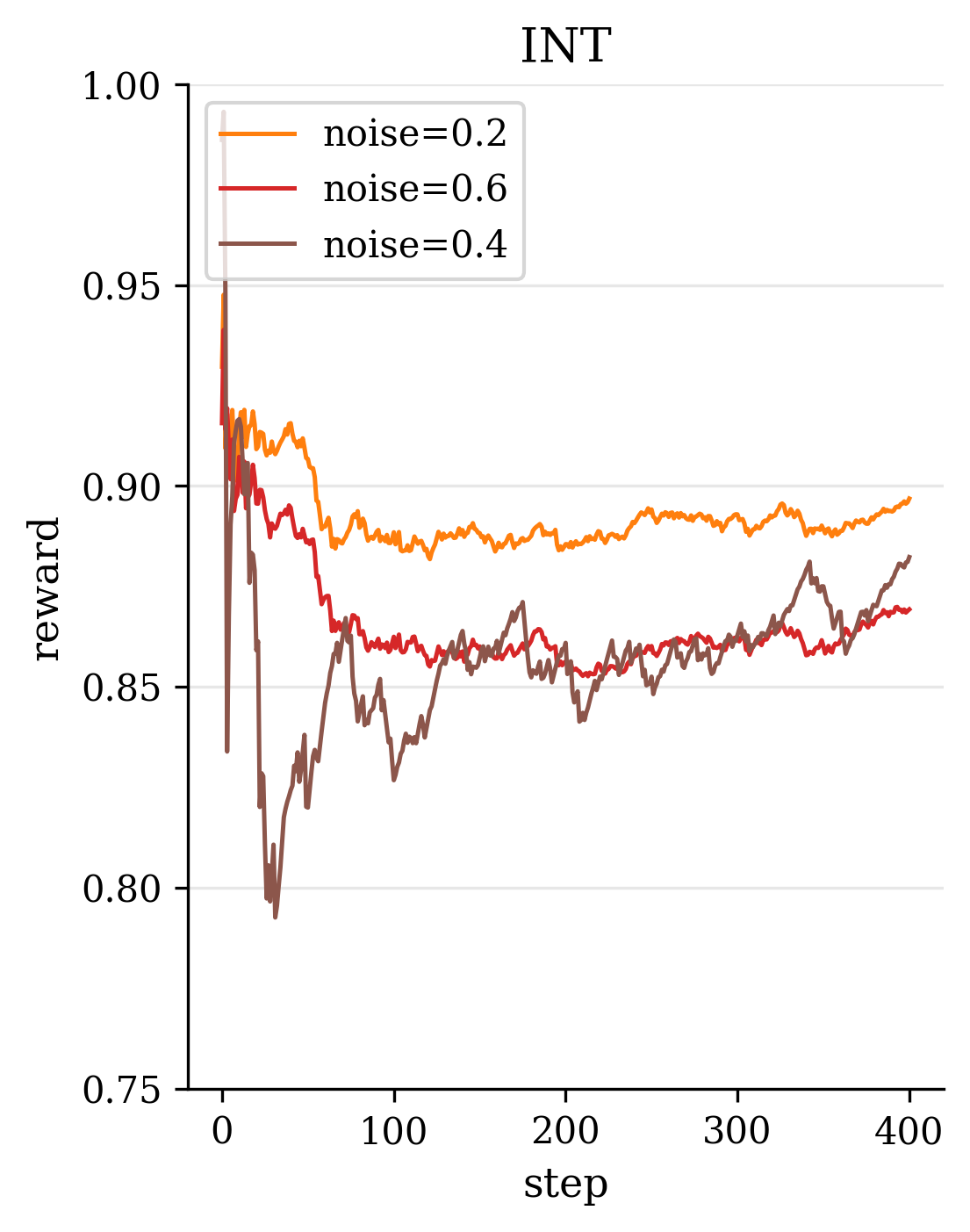}
 
        \label{fig:sub3}
    \end{subfigure}
    \hfill
    \begin{subfigure}{0.24\linewidth}
        \centering
        \includegraphics[width=\linewidth]{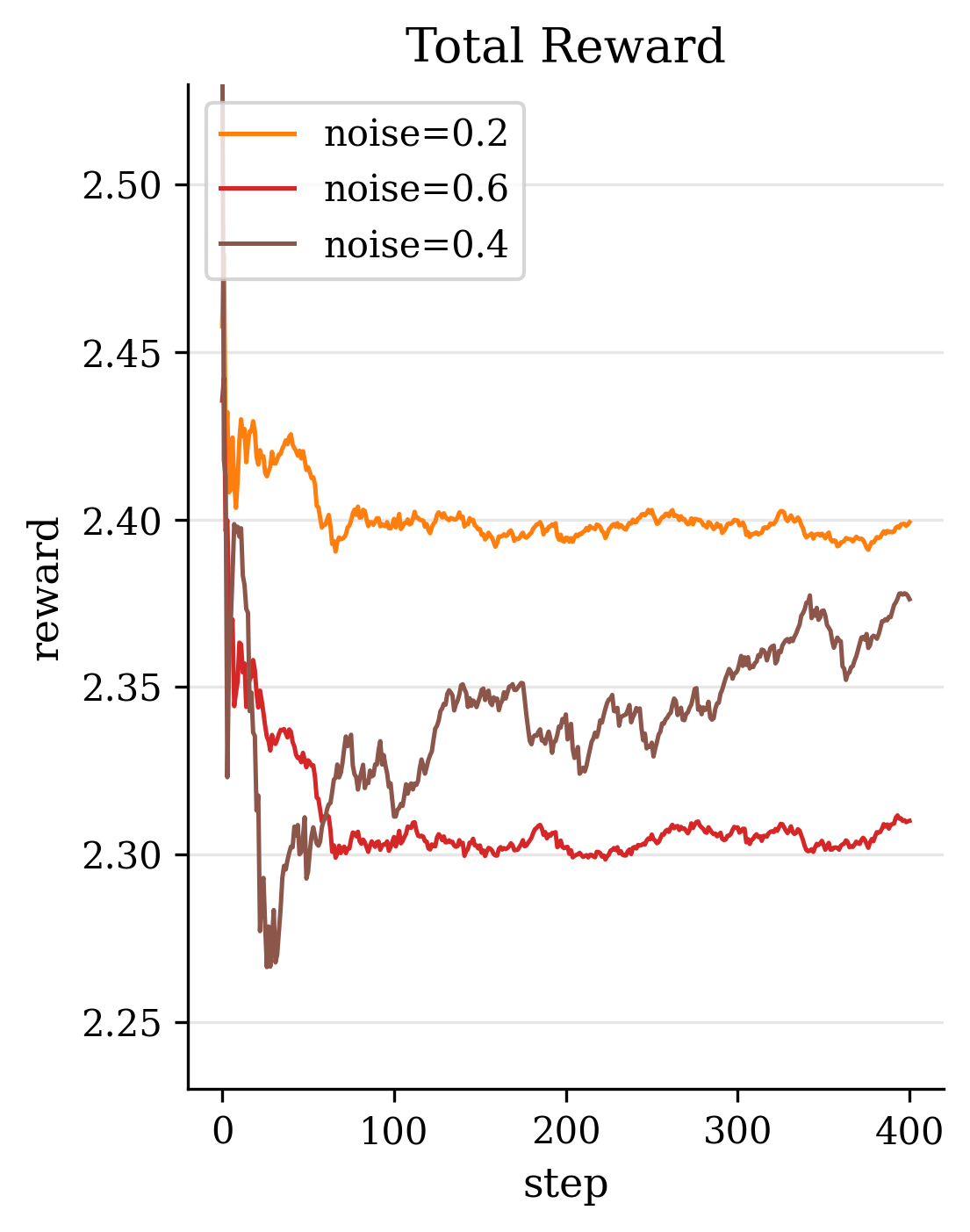}

        \label{fig:sub4}
    \end{subfigure}
    
    \caption{RL ablation experiments on noise level $a$.}
    \label{fig:noise}
\end{figure}

\begin{figure}[t]
    \centering
    \begin{subfigure}{0.24\linewidth}
        \centering
        \includegraphics[width=\linewidth]{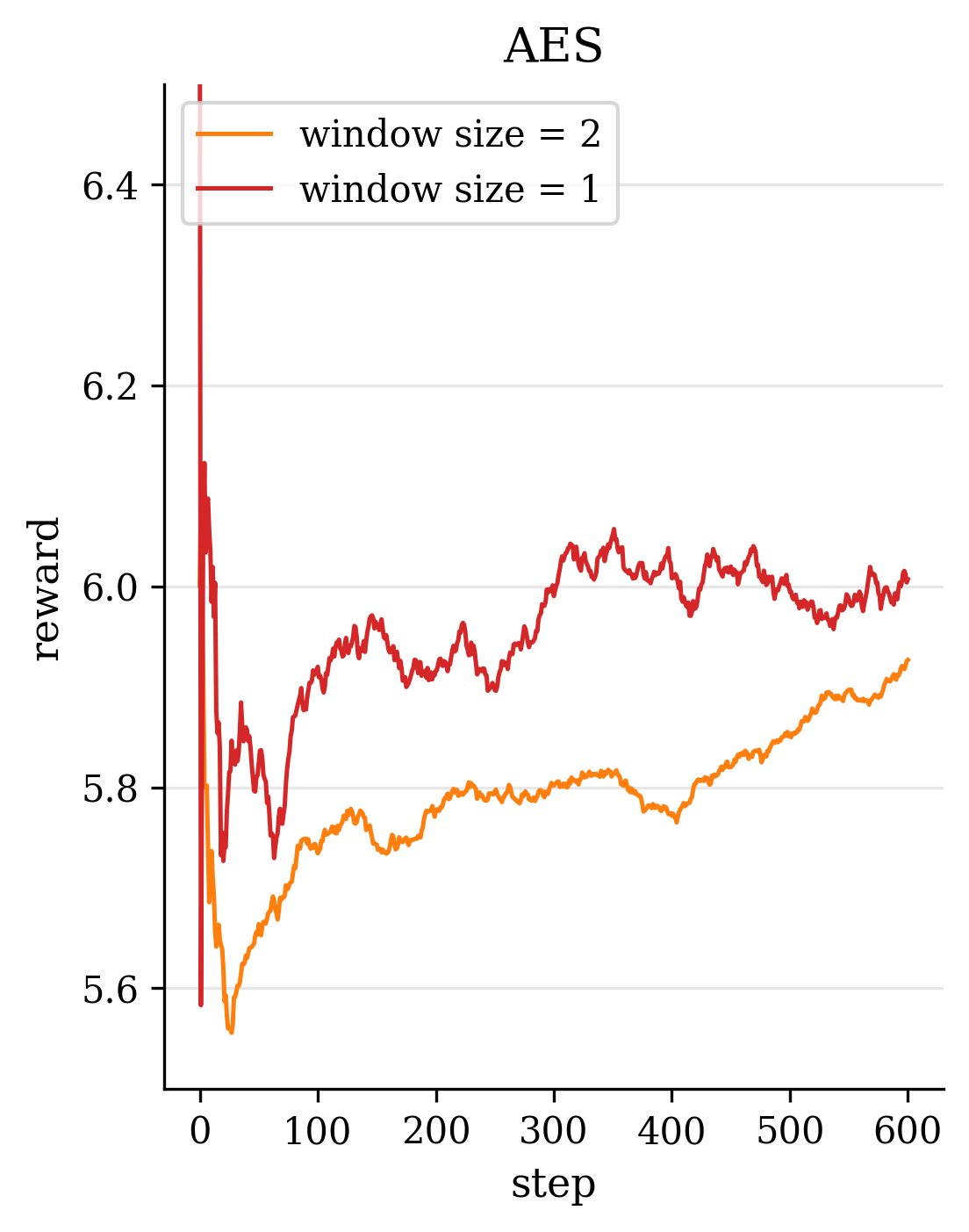}

        \label{fig:sub1}
    \end{subfigure}
    \hfill
    \begin{subfigure}{0.24\linewidth}
        \centering
        \includegraphics[width=\linewidth]{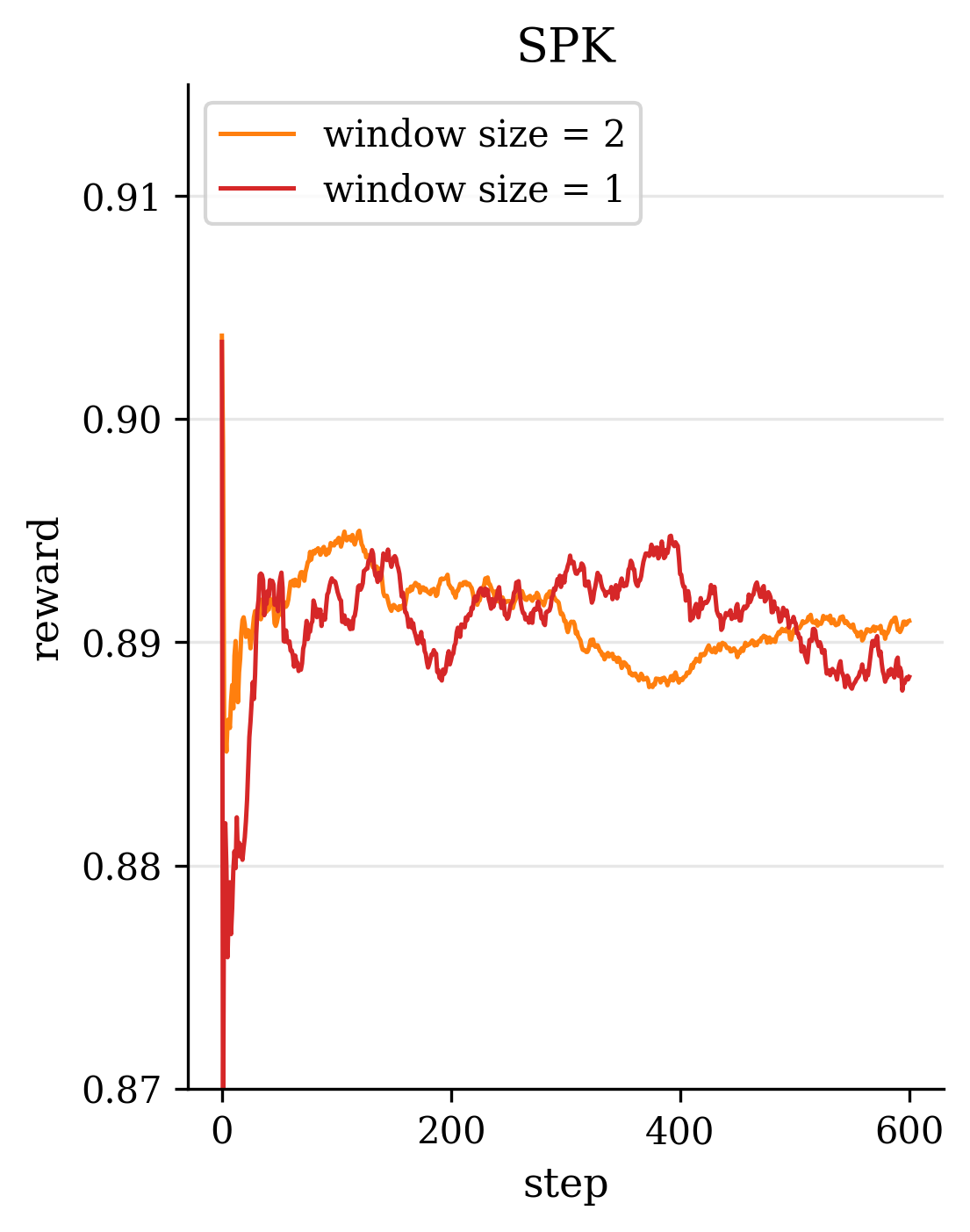}

        \label{fig:sub2}
    \end{subfigure}
    \hfill
    \begin{subfigure}{0.24\linewidth}
        \centering
        \includegraphics[width=\linewidth]{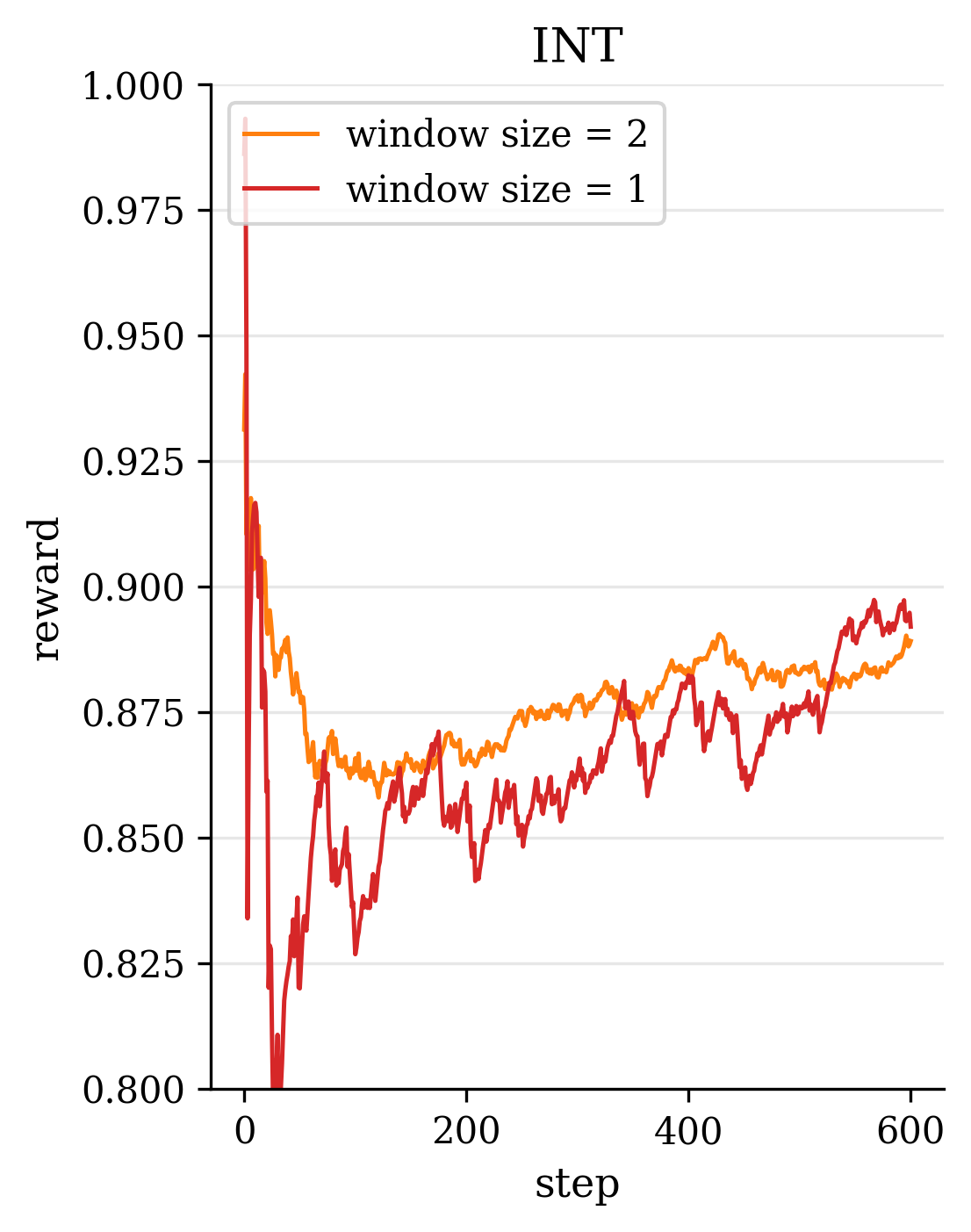}
 
        \label{fig:sub3}
    \end{subfigure}
    \hfill
    \begin{subfigure}{0.24\linewidth}
        \centering
        \includegraphics[width=\linewidth]{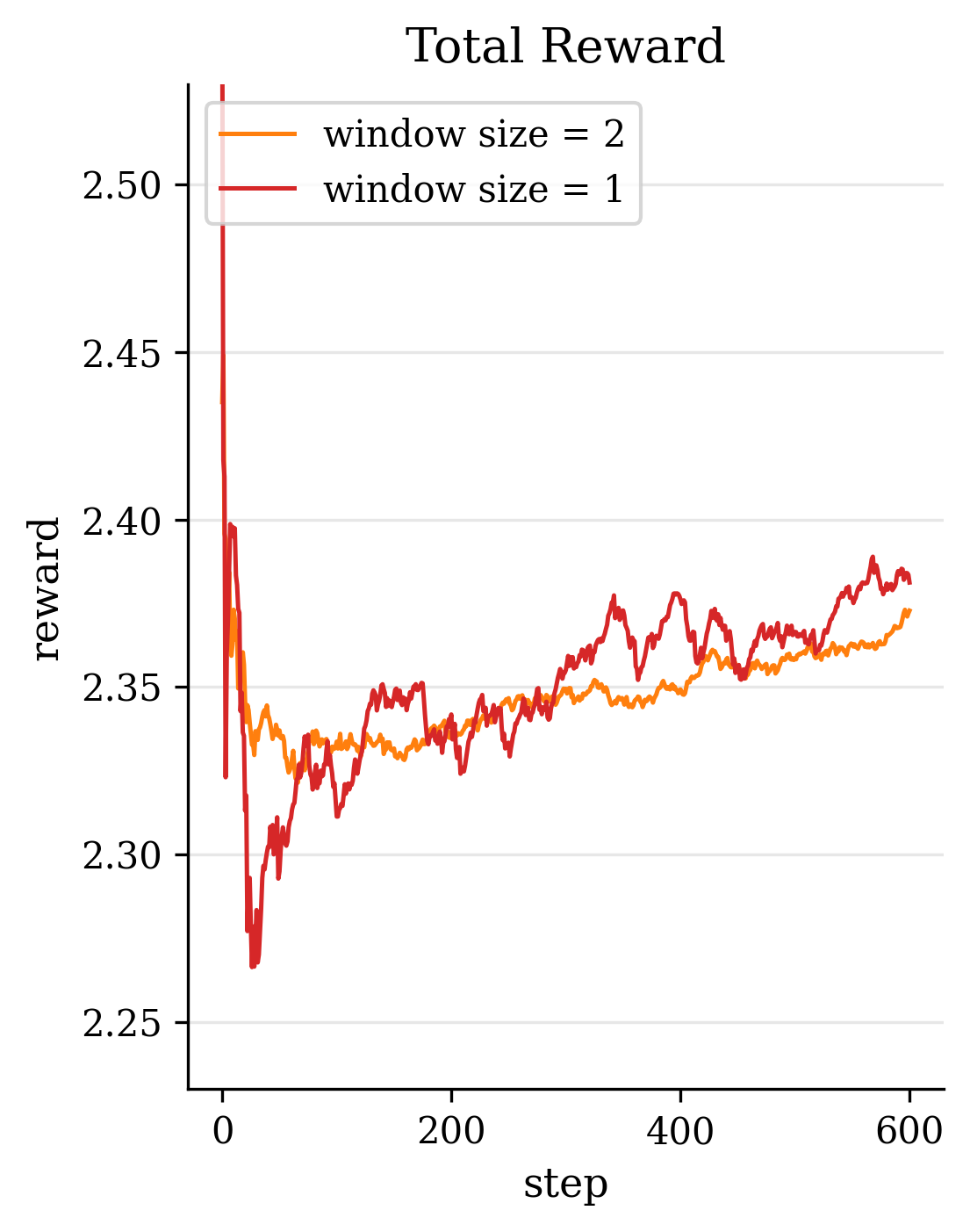}

        \label{fig:sub4}
    \end{subfigure}
    
    \caption{RL ablation experiments on window size.}
    \label{fig:windowsize}
\end{figure}

\subsubsection{Ablation of the RL Setting}

\textbf{Ablation on Noise Level $a$ in GRPO Training.}
In our GRPO-based reinforcement learning stage, the noise level $a$ determines the magnitude of stochastic perturbation injected into the latent trajectory at the selected timestep. This parameter directly affects both the exploration behavior of the policy and the stability of optimization. Empirically, we observe that excessive noise (e.g., $a \geq 0.6$) introduces disproportionately large deviations in the latent representation, leading to disruption of the timbre embedding structure and a noticeable reduction in speaker similarity. The injected perturbation overrides the discriminative timbre cues required for preserving voice identity, thereby degrading SIM scores.

On the other hand, very small noise levels (e.g., $a = 0.2$) also yield suboptimal outcomes. Although the reward curves during RL appear higher due to reduced sampling variance, both objective and subjective metrics indicate degraded naturalness and intelligibility. This behavior is consistent with insufficient exploration: the trajectories remain overly close to the SFT initialization, causing diminished advantage estimates and limiting the policy’s ability to discover reward-improving generation modes.

Across tested values, $a = 0.4$ provides the most balanced behavior. It maintains the structural stability of timbre embeddings while introducing enough stochasticity for effective reward-driven exploration. This configuration yields more stable convergence and consistent improvements across SPK-SIM, aesthetics, and intelligibility metrics, and is therefore adopted as the default setting.

\medskip
\textbf{Ablation on Window Size $S_{\mathrm{window}}$ in GRPO Training.}
We further examine the effect of the temporal window size $S_{\mathrm{window}}$ used during advantage computation. Two configurations, $S_{\mathrm{window}} = 1$ and $S_{\mathrm{window}} = 2$, were compared. A larger window provides additional temporal smoothing and reduces reward variance at early RL iterations; however, it simultaneously weakens the sensitivity of the policy to the reward signal. This reduced sensitivity makes it harder for the RL updates to react to fine-grained differences across the multi-objective reward dimensions, limiting the effectiveness of policy improvement.

In contrast, using $S_{\mathrm{window}} = 1$ yields higher exploration efficiency. Although initial reward curves exhibit greater fluctuation, the policy converges more rapidly and eventually achieves higher final rewards across multiple metrics. Based on these observations, we adopt $S_{\mathrm{window}} = 1$ as the default configuration for our final system.

\textbf{Training Dynamics of Multi-Objective Selective Flow-GRPO.}
During the post-RL optimization stage, the model undergoes a notable redistribution of its policy, which is reflected in a temporary drop in rewards across all metrics. This phenomenon arises because the RL updates deliberately perturb the SFT-initialized policy, forcing it out of its local optimum and enabling exploration. Injecting stochasticity at a single timestep in Selective GRPO reshapes the policy distribution and initially destabilizes phoneme boundaries, pitch trajectories, and timbre embeddings, which in turn degrades intelligibility- and similarity-related rewards. The widened reward variance caused by within-group normalization further accentuates this short-term decline. Importantly, this early reduction does not indicate training failure; rather, it reflects the expected exploratory phase where the model searches for policies that better optimize the composite reward.

As optimization proceeds, the policy gradually adapts to balance the competing objectives. Different reward components (e.g., speaker similarity, intelligibility, aesthetic quality) reach improvement at different rates because they depend on distinct—and sometimes conflicting—acoustic attributes. Enhancing articulation clarity, for instance, may disrupt timbre consistency, while improving spectral smoothness may reduce discriminability required for speaker similarity. Despite these inherent trade-offs, RL enables the model to explore a broader solution space beyond supervised SFT, eventually converging to a policy that yields consistent gains across all dimensions. Empirically, the RL-enhanced model surpasses the pure SFT baseline in timbre preservation, lyric intelligibility, and perceptual naturalness, confirming the effectiveness of multi-objective Selective GRPO for real-world singing voice conversion.

\section{Conclusion}
\label{sec:conclu}
\textbf{YingMusic-SVC} is a robust zero-shot singing voice conversion framework designed for real-world deployment. It integrates multi-stage learning—continuous pre-training, robust supervised fine-tuning, and Flow-GRPO reinforcement learning—with singing-specific inductive biases. By incorporating a singing-trained RVC timbre shifter, an F0-aware timbre adaptor, and an energy-balanced flow matching loss, YingMusic-SVC achieves superior fidelity, expressiveness, and robustness across both clean and accompanied singing scenarios. Experimental results demonstrate state-of-the-art performance compared to other open-source SVC systems, particularly under harmony-contaminated conditions. In future work, we plan to explore adaptive reward modeling for reinforcement learning, multi-style timbre control, and efficient inference optimization for real-time SVC applications.

\newpage

\bibliographystyle{apalike}
\bibliography{refs}






\end{document}